\documentclass[3p,times,twocolumn]{elsarticle}
 \biboptions{comma,sort&compress}
 
\usepackage{graphicx}
\usepackage{amsmath}
\usepackage{here}
\usepackage{ecrc}
\volume{00}

\firstpage{1}

\journalname{Nuclear and Particle Physics Proceedings}
\runauth{Alessandro Ferretti on behalf of the ALICE Collaboration}

\jid{nppp}
\jnltitlelogo{Nuclear and Particle Physics Proceedings}
\usepackage{amssymb}
\usepackage[figuresright]{rotating}

\begin{document}

\begin{frontmatter}

%%
%%%%%%%%%%%%%%%%%%%%%%%%%%%%%%%%%%%%%%%%%%%%%%%%%
%\begin{document}
\title{
%$\la g^3f_{abc} G^aG^bG^c\ra$
%$\la g^3 f_{abc}G^3\ra$ 
% 
%$\alpha_s $, $\la \alpha_sG^2\ra$, $\overline{m}_{c,b}$ and $f_{B_c}$
%from  relativistic heavy quark sum rules$^*$} 
ALICE upgrades for Run 4 and Run 5}
 
 %\cortext[cor0]{Mini-Review talk presented at QCD20 - 35 years later, 23th International Conference in QCD %(27-30/10/2020,
 % Montpellier - FR). }

\author[1]{Alessandro Ferretti}
\ead{ferretti@to.infn.it}
\author[2]{on behalf of the ALICE Collaboration}
 %\corref{cor1} \andnewline
   \address[label1]{Università di Torino, Dipartimento di Fisica, via Giuria 1, 10125 Torino, Italy \\
and\\
INFN - Sezione di Torino, via Giuria 1, 10125 Torino, Italy
}

%\cortext[cor1]{ICTP-Trieste  researcher consultant for Madagascar.}

\pagestyle{myheadings}
\markright{ }
\begin{abstract}
\noindent

In view of Run 4 at the LHC, presently scheduled from 2029 onwards, ALICE is pursuing several upgrades to further extend its physics reach.
In order to improve heavy-flavor hadron and dielectron measurements which rely on secondary vertexing, a reduction of the material budget of the innermost layers of the Inner Tracking System is needed. This can be achieved with bent pixel sensors, arranged in half-cylinder shapes with integrated power lines and data buses, allowing to get rid of most of the supporting structure and of the water cooling.
Moreover, a new Forward Calorimeter (FoCal), covering pseudorapidities of $3.2<\eta<5.8$, has been proposed
to measure small-x (down to $10^{-6}$) gluon distributions via prompt photon production. The FoCal will be composed of a highly granular Si+W electromagnetic calorimeter combined with a conventional sampling hadronic calorimeter, and will significantly enhance the scope of ALICE for inclusive and correlation measurements with mesons, photons, and jets.
For Run 5 and beyond, the ALICE 3 project has been proposed. It consists of a novel compact detector based on monolithic silicon sensors, with ultra-thin layers near the vertex, high readout rate capabilities, superb pointing resolution, excellent tracking and particle identification over a large acceptance. Such a detector enables a rich physics program, ranging from electromagnetic probes at ultra-low transverse momenta to precision physics in the charm and beauty sector. For particle identification, a sequence of detector systems is foreseen: a combination of a Time-Of-Flight system, a Ring-Imaging Cherenkov detector, an electromagnetic calorimeter, a muon identifier, and a dedicated forward detector for ultra-soft photons.
In these proceedings, the upgrade plans will be shown together with the status of R\&D on ITS3 and FoCal and with the concepts and requirements of ALICE 3.

%We briefly report the modern status of heavy quark sum rules (HQSR) based on stability criteria by emphasizing the recent progresses for determining the QCD parameters ($\alpha_s$, $m_{c,b}$ and gluon condensates) where their correlations have been taken into account.  
 
%% keywords
%\ccode{Pac numbers: 11.55.Hx, 12.38.Lg, 13.20-Gd, 14.65.Dw, 14.65.Fy, 14.70.Dj}  
\end{abstract}
\end{frontmatter}
%%%%%%%%%%%%%%%%%%%%%%%%%%%%%%%%%%
%\end{document}
%%%%%%%%%%%%%%%%%%%%%%%%%%%%%%%%%%
%\vspace*{-1.5cm}
\section{Introduction}
%\vspace*{-0.25cm}
 %\nin
%%%%%%%%%%%%%%%%%%%%%%%%%%%%%%%%%%%

ALICE (A Large Ion Collider Experiment) is one of the experiments at the CERN Large Hadron Collider (LHC) and it is mainly devoted to heavy-ion physics \cite{alice1}. Its main purpose is the detailed characterization of strongly interacting Quark-Gluon Plasma (QGP): to this end, ALICE is designed to carry out comprehensive studies of hadrons, electrons, muons, heavy-flavors hadrons, photons and jets produced in heavy-ion collisions.

During the LHC Long Shutdown 2 (LS2, Dec. 2018- May 2022) ALICE installed a major upgrade which included a new Inner Tracking System (ITS2) made of seven concentric layers of Monolithic Active Pixel Sensors (MAPS) with an ALPIDE design \cite{alpide}, the replacement of TPC readout wire chambers with GEMs (Gas Electron Multiplier), the addition of a silicon-based Muon Forward Tracker (MFT) in front of the Muon Spectrometer, a Fast Interaction Trigger (FIT) system and a new integrated Online-Offline system (O$^2$) for data acquisition, calibration, reconstruction and storage \cite{alice2}.
In view of Run 4 (scheduled in 2029-2032) the ALICE Collaboration is working on two new major upgrades, which will involve the replacement of the inner layers of ITS2 with new, low-material budget, bent silicon detectors (ITS3) and a new, high-granularity Forward Calorimeter (FoCal). In addition, in view of Run 5 (2035-onwards), the collaboration has proposed a next-generation detector (ALICE 3) with increased rate capabilities, improved vertexing and tracking over a wide momentum range and rapidity coverage. Details of these upgrades will be given in the following.

\section{Inner Tracking System (ITS3) for Run 4}
The Inner Tracking System was completely replaced during LS2 in order to enhance vertex and tracking precision. In view of Run~4, ALICE plans to further improve the ITS performances by replacing the 3 innermost layers \cite{its3}. The new Inner Barrel will be closer to the Interaction Point (IP) to improve vertex resolution: this will require a new beam pipe with smaller inner radius (from 18.2~mm to 16~mm) and reduced thickness (from 800 to 500~$\mu$m). The new silicon layers will be realized with 65~nm CMOS technology, which features a power consumption below 20~mW/cm$^2$ allowing the removal of the water cooling. Power lines and data buses will be integrated directly on the sensor, allowing the removal of Printed Circuit Boards. 

When thinned down to 20-40~$\mu$m, the sensors become flexible and can be bent. Moreover, up to 300~mm large sensors should be feasible thanks to the use of stitching technology (under investigation, a dedicated R\&D effort is ongoing). This means that large sensors can be bent and cut in the shape of half-cylinders with the desired radius. The increased stiffness of bent silicon wafers, together with the use of air cooling, will permit to replace the support structure with low-density carbon foam ribs. Six half-cylindrical sensors (1 for each half layer) are foreseen (see fig.~1). The material budget will be lowered from the present mean value of 0.35\% to below 0.05\% of a radiation length, and the removal of the support structure will lead to very homogeneous material distribution, lowering associated systematic unvertainties to a negligible level. 
\begin{figure}[h]
\centering
\label{its3}
\includegraphics[width=0.5\textwidth]{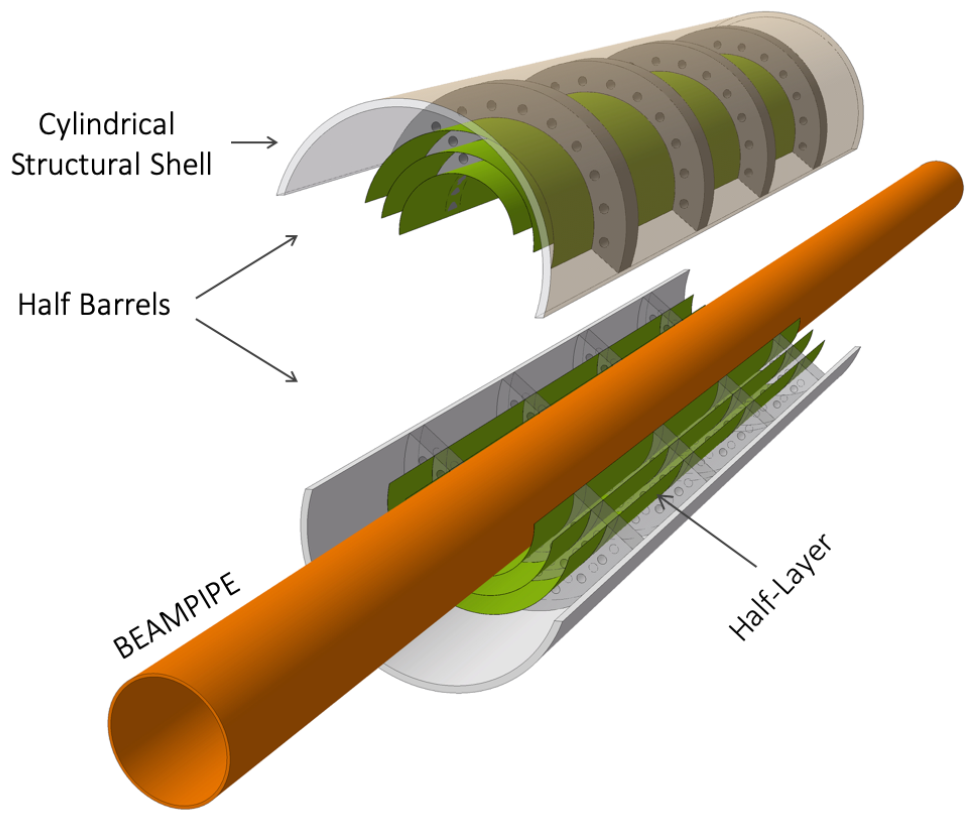}
\caption{Layout of the ITS3 Inner Barrel. The figure shows the two half-barrels mounted around the beampipe}
\end{figure}

After mechanical and electrical characterization in the laboratory, full performance characterization of 50 $\mu$m thick ALPIDE sensors bent to the ITS3 radius has been carried out in 2021 with a beam test \cite{testbent}. The measured inefficiency was generally below $10^{-4}$, regardless of the inclination and position of the impinging beam with respect to the sensor surface.
Simulations of the ITS3 expected physics performance show an enhancement of pointing resolution by a
factor of 2 over all momenta, and an increase of tracking efficiency for low-$p_\mathrm{T}$ particles: with ITS2, tracking efficiency falls below 50\% for transverse momentum values smaller than 80~Mev/$c$, while with ITS3 the 50\% threshold will be lowered to less than 70~MeV/$c$. \cite{its3}.

\section{Forward Calorimeter for Run 4}
The Forward Calorimeter (FoCal)\cite{focal} is an ALICE upgrade which will add new capabilities to explore the small-x parton structure of nucleons and nuclei down to Bjorken-x values of $\sim10^{-6}$, thanks to high-precision inclusive measurements of direct photons and jets and coincident photon-jet and jet-jet measurements. It will consist of a highly granular Si+W electromagnetic calorimeter combined with a conventional sampling hadronic calorimeter, placed 7~m from the IP around the beam pipe and covering a pseudorapidity interval $3.2<\eta<5.8$ (fig.2).

The EM calorimeter (FoCal-E) main challenge is the $\gamma /\pi^0$ separation at high energy. In the FoCal position, transverse separation between photons originated from one neutral pion decay is of the order of 2~mm: this requires an absorber material with a small Molière radius such as tungsten. Concerning the position resolution, simulations show that a readout granularity as fine as 0.3~mm is needed to provide crucial information for $\pi^0$ identification.

The required calorimeter total radiation length is of the order of 20 $X_0$, which can be achieved with 20 tungsten layers 3.5~mm thick (about 1~$X_0$ each): however, instrumenting 20 layers with pixel sensors would lead to high costs of the readout system and to large readout data volumes. After extensive simulation studies, the optimal solution has been found to be a combination of 18 low-granularity (1~cm) silicon pad layers, which will provide the shower profile and the total energy, and 2 high granularity pixel readout layers, based on ALPIDE sensors used for the ITS2 and placed after the $5^\mathrm{th}$ and $10^\mathrm{th}$ absorber layers, giving the position resolution needed to resolve overlapping showers. The ALPIDE pixel size is 30x30 $\mu$m, i.e. smaller than needed, but its choice permits to reuse a technology directly compatible with the ALICE environment and to obtain pseudoanalog information from the density of fired pixels.
Overall, this hybrid approach will provide the required readout granularity while keeping to a minimum both costs and data volumes. A first test of the FoCal-E, equipped with 18 pad layers, has been carried out at the CERN PS in 2022 showing a clear separation of the MIP peak and pedestal noise: further tests are planned at the SPS in Autumn, 2022.
\begin{figure}[h]
\centering
\label{fig2}
\includegraphics[width=0.5\textwidth]{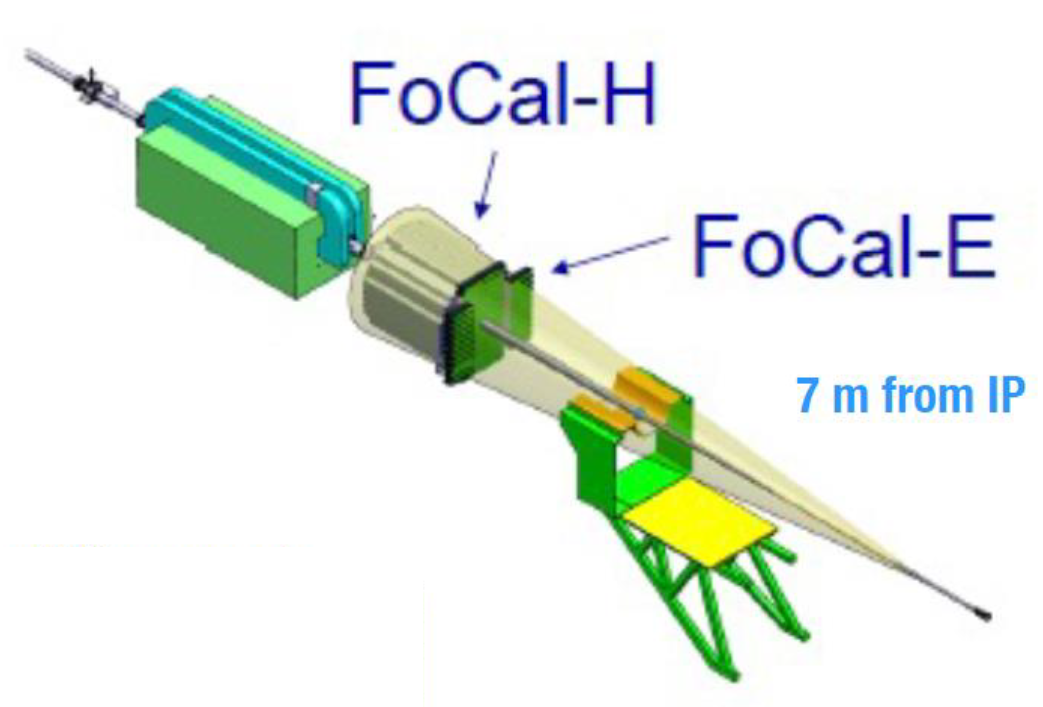}
\caption{Layout of the FoCal detector and its position along the beam pipe.}
\end{figure}

The electromagnetic calorimeter of the FoCal will be complemented with a hadronic calorimeter (FoCal-H) for photon isolation and jet measurements. Here, a rather conventional metal-scintillator spaghetti-design was chosen. A first prototype has been realized with 1440 capillary copper tubes 550 mm long, aligned along the beam line (outer diameter 2.5~mm. inner diameter 1.2~mm) and arranged in a 36x40 matrix (overall dimensions 95x95x550 $\mathrm{mm}^3$) The tubes are filled with scintillating fibers (diameter 1.0~mm) read out by silicon photomultipliers. This prototype was tested with the SPS beam up to an energy of 120 GeV: test results showed a good consistency between MonteCarlo simulation and real data. A second prototype is currently under preparation: it will be based on 9 copper tube modules  65x58x1100 $\mathrm{mm}^3$ each, arranged in a 3x3 matrix.

\section{ALICE 3}
Thanks to the upgrades described in previous sections, ALICE will be able to extend considerably its physics reach. However, we already know that some important questions will have only partial answers. In order to collect data toward this goal a novel experimental approach will be needed, especially in the field of heavy-flavor and electromagnetic radiation studies.  Systematic measurements of heavy-flavor hadrons,  with particular attention to particles featuring multiple charm quarks or beauty quarks, would allow to disentangle effects due to the transport properties in the QGP from the ones arising from hadronization mechanisms. A precise measurement of dileptons would shed light on the evolution of the QGP and allow to explore the mixing between $\rho$ and a$_1$ mesons, which would represent a fundamental confirmation of chiral symmetry restoration in the medium. Furthermore, higher rate capabilities would allow to access rare events which are key topics in the study of the QGP, hadronic structure and nuclear physics, such as the production of nuclei, hypernuclei and possibly the as-yet undiscovered supernuclei. Finally, a wider acceptance associated with powerful particle identification capabilities at low-$p_\mathrm{T}$ would allow to better study hadron interaction potentials via two-particle correlations of charmed hadrons.

In order to achieve these goals, a new detector has been proposed, to be placed in the ALICE cavern during LHC Long Shutdown 3 and able to take data from the start of LHC Run~5 (currently foreseen in 2035). Its main characteristics are an unprecedented pointing resolution to measure secondary vertices (crucial for heavy-flavor identification), ample PID capabilities coupled with large rapidity acceptance and a high data-taking rate (see fig.~3).

\begin{figure}[h]
\centering
\label{fig3}
\includegraphics[width=0.5\textwidth]{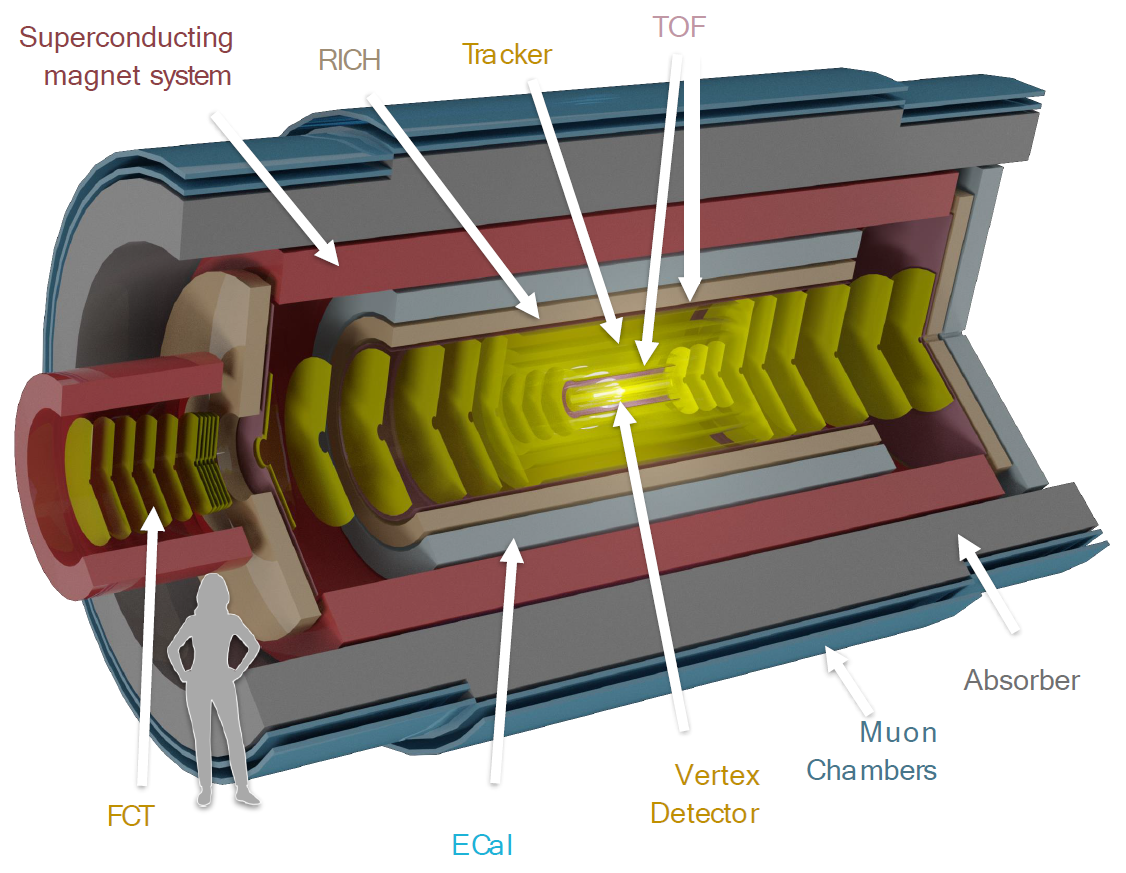}
\caption{Layout of the ALICE3 experiment.}
\end{figure}

The core of the ALICE 3 system is a full silicon tracking device, made of 11 barrel layers completed by 2x12 endcap disks. The inner 3 layers and 2x3 disks make up the Vertex Detector. Its pseudorapidity acceptance interval is  $|\eta|<4$, and its radial extension ranges from 80~cm down to 0.5~cm. In order to meet the requirements of pointing resolution, the first layer needs to be very close to the IP and its radiation length should be about $0.1\%~X/X_0$, with a position resolution of $2.5~\mu$m.

The Time-Of-Flight (TOF) system is made of two cylindrical  layers with a time resolution of 20~ps. The larger layer, with a 85~cm radius, will allow to separate electrons from hadrons up to $p_\mathrm{T}=500~\mathrm{MeV}/c$ and $\pi/$K separation up to 2~GeV/$c$, while an inner layer will be placed between the tracking layers to extend the reach to lower $p_\mathrm{T}$. A cell pitch of the order of 1~mm is needed, and the single layer $X/X_0$ ratio must be kept below 3\%.

The TOF outer layer is surrounded by a cylindrical Ring Imaging CHerenkov (RICH)  detector, enabling $e/\pi$ separation up to 2~GeV/$c$ and to distinguish protons from $e, \pi$ and $K$ up to 14~GeV/$c$.
The RICH detector is enclosed inside an electromagnetic calorimeter covering the barrel acceptance and one forward direction, enabling measurements of photon-jet correlations and radiative decays of charmed mesons.
All these detectors will be hosted inside a superconducting magnet, providing the magnetic field needed to measure momenta from track curvature: a solenoidal field of $B=2$~T is sufficient for a relative $p_\mathrm{T}$ resolution of $~0.6\%$ at mid-rapidity.

For muon identification a cylindrical steel absorber of $\sim 70$~cm thick, surrounding the magnet, shields two layers of muon detectors, used to identify muons  by matching hits in the muon detectors with tracks from the inner tracker. Absorber thickness is optimized for the reconstruction efficiency of J/$\psi$ emitted at rest at zero rapidity, i.e. muons with a transverse momentum as low as 1.5~GeV/$c$.

Finally, a Forward Conversion Tracker (FCT) is foreseen to measure photons down to transverse
momenta of $\sim 2$ MeV/$c$, which are only accessible by exploiting the Lorentz boost in the forward
direction resulting in photon energies down to $\mathrm{E}_\gamma \sim 50$ MeV at $\eta= 4$.

\subsection{Main R\&D challenges}
The ALICE 3 project will require a dedicated R\&D effort in order to ovecome non-trivial challenges. 
The innermost Vertex Tracker layer radius (5~mm) is smaller than the LHC beam transverse radius during the injection phase (10~mm): this means that the Vertex Tracker has to be placed inside the beam pipe, and a mechanism has to be devised in order to retract the sensors out of harm's way during the LHC injection phase.

Moreover, a position resolution of the order of 1~$\mu$m will need a pixel pitch smaller than 10 $\mu$m.
The outer tracker presents a large instrumented area (of the order of tens of square meters) which requires the development of cost-effective sensors, and to keep its material budget as low as 1\% $X/X_0$ per layer will imply the realization of low-weight supports and services.

Concerning the TOF system, the need for a $<20$~ps time resolution on the system level implies advances both on the sensors and microelectronics. The goal is to develop monolithic sensors in a CMOS technology with the required time resolution, while keeping the cost reasonably low.
\section{Conclusions}
Progress in the understanding of ultrarelativistic heavy-ion collisions is strictly dependent on improvements in the physics measurements, which in turn calls for upgrades in the experimental apparatus. The ALICE collaboration has formulated a roadmap to an exciting future physics program supported by strategic detector developments in order to make it suitable for exploration of the many important questions that still lay unanswered in this knowledge field.
%\vfill\eject
%%%%%%%%%%%%%%%

\end{document}